# On the Mechanism of the Spin State Transition of (Pr$_{1-y}$Sm$_y$)$_{1-x}$Ca$_x$CoO$_3$

Toshiaki Fujita, Shohei Kawabata and Masatoshi Sato[*]

*Department of Physics, Nagoya University, Furo-cho, Chikusa-ku, Nagoya 464-8602*

Nobuyuki Kurita, Masato Hedo and Yoshiya Uwatoko

*Institute for Solid State Physics, The University of Tokyo, 5-1-5 Kashiwanoha, Kashiwa, Chiba, 277-8581*

Transport, thermal and magnetic measurements have been carried out on (Pr$_{1-y}$Sm$_y$)$_{1-x}$Ca$_x$CoO$_3$. The system exhibits a structural phase transition accompanied by the spin state change from the intermediate spin (IS) state to the low spin (LS) state with decreasing temperature $T$. We have constructed a $T$-$y$ phase diagram for $x$=0.3 and $T$-$x$ ones for $y$=0.2 and 0.3. By analyzing their magnetic susceptibilities, the number of Co ions excited to IS state (or the electron number in the $e_g$ orbitals), $n_{IS}$, are roughly estimated. With increasing $y$ or with decreasing $x$, $n_{IS}$ decreases, and the phase transition changes gradually to the (IS→LS) crossover-like one. We discuss on the possible role of the Pr atoms in realizing the transition.

## 1. Introduction

Co oxides, which have the linkage of $CoO_6$ octahedra, have attracted much interest, because they exhibit various notable physical characteristics. The superconductivity found in $Na_zCoO_2 \cdot yH_2O$ ($z \sim 0.3$ and $y \sim 1.3$) is one of the examples of such characteristics.[1-4] The spin state change often observed for $Co^{3+}$ ions in perovskite systems $RCoO_3$ (R= various rare earth elements and Y) is another example. They exhibits a spin state change from the low spin (LS; spin $s=0$; $t_{2g}^6$) ground state to the intermediate spin (IS; $s=1$; $t_{2g}^5 e_g^1$) or the high spin (HS; $s=2$; $t_{2g}^4 e_g^2$) state with increasing temperature $T$.[5-16] The existence of the spin state change indicates that the difference of the electronic energies, $\delta E$, between these states is rather small. Therefore, we can control the physical properties of Co oxides by controlling the value of $\delta E$ by various methods. For example, the change of the ionic radius $r_R$ of $R^{3+}$ clearly affects the value of $\delta E$, because the pseudo cubic crystal field splitting $\Delta_c$ between 3-fold $t_{2g}$ and 2-fold $e_g$ orbitals depends on $r_R$ through the volume change of the $CoO_6$ octahedra. The Co-Co transfer energy $t$, which seems to be important to determine $\delta E$,[17] also depends on $r$ through the change of Co-O-Co bond angle induced by the change of the tolerance factor $(r_R+r_O)/\sqrt{2}(r_{Co}+r_O)$, where the $r_M$ (M=R, O and Co) is the ionic radius of M. The increasing tendency of the temperature of the spin state change with decreasing R-ion size is explained by these facts.[15-17] Typical results of the studies on the relationship between the spin state of $Co^{3+}$ ions and the local structures can be found in our reports, too.[17-20]

In the consideration of $\delta E$ of $R_{1-x}A_xCoO_3$ (A= Ba, Sr and Ca), effects of the hole doping should also be considered.[21-27] The substitution of $R^{3+}$ with $A^{2+}$ affects the effective values of $\delta E$ through the introduction of the hole-carriers and by changing the average ionic radius of $R_{1-x}A_x$. For the relatively large atom species, for example, R=La and/or A=Ba and Sr, the ferromagnetic transition is often observed, which is due to the enhancement of the double exchange interaction caused by the decrease of $\delta E$ or increase of the electron number in the $e_g$ orbitals.

For the samples of $(Pr_{1-y}R'_y)_{1-x}Ca_xCoO_3$ (R'=rare earth elements and Y), a phase transition which is accompanied by the change from the IS to LS state of Co ions with decreasing $T$, is observed.[17,28] In the previous paper, we studied the transition in detail for $(Pr_{1-y}R'_y)_{1-x}Ca_xCoO_3$ ($0.0 \leq x \leq 0.5$; $0.0 \leq y \leq 0.2$) to extract information by what factors and how the spin state change is governed, where various kinds of studies, transport and magnetic measurements as well as the structure analyses at ambient pressure and high pressure were carried out.[17] We found that the sudden increase of the tilting angle of the $CoO_6$ octahedra at the transition temperature $T_s$ with decreasing $T$ mainly stabilizes the LS state through the reduction of the Co-Co transfer energy $t$. The local arrangements of the ligand atoms seem to be important to understand the structural transition or the spin state change. For example, the substitution of Pr with elements whose ionic radii are smaller than $Pr^{3+}$, enhances $T_s$ through the increase of $\Delta_c$ and through the decrease of $t$ (induced by the increase of the tilting angle). However, there remains a question why the transition takes place only in systems, which contain Pr and Ca.[17, 21] Then, we have carried out further studies by various methods, transport, thermal and magnetic measurements for samples of $(Pr_{1-y}Sm_y)_{1-x}Ca_xCoO_3$ (0.0≤$x$≤0.5, 0.0≤$y$≤1.0) to clarify details of the transition.

In the present paper, the $T$-$y$ and $T$-$x$ phase diagrams are constructed and the mechanism of the spin state transition of the systems is discussed.

## 2. Experiments

Polycrystalline samples of $(Pr_{1-y}Sm_y)_{1-x}Ca_xCoO_3$ (0.0≤$x$≤0.5; 0.0≤$y$≤1.0) were prepared by the following method. Mixtures of $Pr_6O_{11}$, $Sm_2O_3$, $CaCO_3$ and $CoC_2O_4 \cdot 2H_2O$ with proper molar ratios were ground, and pressed into pellets. The pellets were sintered at 1200 ºC for 24 h under flowing oxygen and cooled at a rate of 100 K/h. Samples thus obtained were annealed in high pressure oxygen atmosphere ($p$~60 atm) at 600 ºC for 2 days. In powder X-ray diffraction patterns taken with Fe$K\alpha$ radiation, no impurity phases were detected. Details of the sample characterization can be found in ref. 17.

Electrical resistivities ρ were measured by the standard four-terminal method by using an ac-resistance bridge on heating from 4.2 K to 300 K. Magnetic susceptibilities χ were measured by using a Quantum Design SQUID magnetometer under the magnetic field $H$ of 0.1 T in the temperature range of 5-350 K. The thermoelectric powers $S$ were measured by the DC method, where the typical temperature difference between both ends of the sample was about 1 K. The specific heats $C$ were measured by a thermal relaxation method in the temperature range of 2-300 K using of the Physical Property Measurements System (PPMS, Quantum Design).

## 3. Experimental Results and Discussion

In Fig.1, the unit cell volumes of $(Pr_{1-y}Sm_y)_{0.7}Ca_{0.3}CoO_3$ determined by the X-ray diffraction at room temperature are plotted against $y$. In the determination of the lattice parameters, we have assumed that the systems are orthorhombic and the volume is estimated for the cell described by ~√2$a_p$×2$a_p$×√2$a_p$ (space group *Pnma*); $a_p$ being the lattice constant of the cubic perovskite cell, though it was difficult in the present experimental resolution to distinguish whether the samples with $y$≠0 is strictly orthorhombic or not. (Our previous neutron and X-ray Rietveld analyses indicate the sample with $y=0.0$ has the orthorhombic unit cell.[17]) For the present system, the volume decreases with $y$, because ionic radius of $Sm^{3+}$ is smaller than that of $Pr^{3+}$.[29]

Figure 2 shows the electrical resistivities $\rho$ of the $(Pr_{1-y}Sm_y)_{0.7}Ca_{0.3}CoO_3$ samples. The samples with $y = 0.0$ and $0.1$ do not exhibit the transition. For these samples, the resistivities $\rho$ increase gradually with decreasing $T$ down to their Curie temperature (~65 K) and then, exhibit the metallic $T$ dependence below the temperature. The phase transition is induced by the 20 % doping of Sm to the Pr sites and the transition temperature $T_s$ increases with increasing $y$. The resistivity $\rho$ above $T_s$ also increases with increasing $y$. However, the transition is gradually smeared out as $y$ approaches unity.

In Fig. 3, the magnetic susceptibilities $\chi$ are plotted against $T$ for the samples of $(Pr_{1-y}Sm_y)_{0.7}Ca_{0.3}CoO_3$. The data were taken with the condition of the zero field cooling under the external magnetic field of $H = 0.1$ T. For $y \leq 0.1$, no structural transition is observed. With increasing $y$ through 0.2, the transition appears, and $\chi$ decreases abruptly at $T_s$ with decreasing $T$ for the relatively small values of $y$, indicating that the system undergoes the first order transition, in which the spin state change from the IS to LS with decreasing $T$, is involved. As $y$ approaches unity, the transition is smeared out. These results correspond well to those of the $\rho$ measurements shown in Fig. 2.

In Fig. 4, the thermoelectric powers $S$ for the samples of $(Pr_{1-y}Sm_y)_{0.7}Ca_{0.3}CoO_3$ are shown against $T$. The sign of $S$ is positive in the whole $T$ region. For $y \geq 0.2$, $S$ increase abruptly at $T_s$ with decreasing $T$ for the relatively small values of $y$ and again the anomaly is smeared out as $y$ approaches unity.

The measurements of the specific heat $C$ have also been performed on $(Pr_{1-y}Sm_y)_{0.7}Ca_{0.3}CoO_3$. The results for $y = 0.3 \sim 1.0$ are shown between 2 K and 300 K in Fig. 5. As can be expected from the data shown in Figs. 2-4, $C$ exhibits the sharp anomaly, too, at $T_s$ for the relatively small values of $y$ and the anomaly is smeared out as $y$ approaches unity. It is found that the $C$ values below $T_s$ are nearly equal for all the samples shown here, while the $C$ values above $T_s$ decreases with increasing $y$. Because the atomic masses and the chemical properties of $Pr^{3+}$ and $Sm^{3+}$ are similar, this $y$ dependence may be considered to be related to the $y$ dependence of the spin state excitation of Co ions.

The specific heat $C$ divided by $T$ is plotted against $T$ in the inset of Fig. 5 for the samples of $(Pr_{1-y}Sm_y)_{0.7}Ca_{0.3}CoO_3$ with $y=0.3$, 0.6, 0.8 and 1.0. Below 10 K, $C/T$ increases with decreasing $T$ down to 2 K. It may originate from the spins of randomly distributed $Co^{4+}$ ions in the LS state ($s=1/2$). In this respect, Tsubouchi *et al.* observed a broad peak of $C/T$ at around 3 K for a sample of $Pr_{0.5}Ca_{0.5}CoO_3$[28], and attributed it to the possible the spin ordering of $Co^{4+}$ ions. Here, we note that the magnetic susceptibility of the Co spins in low temperature phase exhibits the Curie-Weiss type $T$ dependence, $\chi_{Co}^{spin} \propto 1/(T-T_C)$ with $T_C \sim 15$ K as shown latter.

In Fig. 6, $T_s$ values of the $(Pr_{1-y}Sm_y)_{0.7}Ca_{0.3}CoO_3$ are shown schematically against $y$, where the gradual change from the first order transition to the crossover-like one is shown by the $y$ dependent broadening of the transition region. Why is the transition smeared out in the region of $y$ close to unity? We think that the answer can be found in the fact that the number the electrons $n_{IS}$ excited in the $e_g$ orbitals becomes small: Because the average ionic radius of $Pr_{1-y}Sm_y$ decreases with increasing $y$, the volume of the $CoO_6$ decreases, causing the increase of $\Delta_c$ and $\delta E$ and therefore the decrease of $n_{IS}$. Due to this decrease of $n_{IS}$, the IS→LS transition gradually lose the cooperative character and the crossover-like nature appears, as $y$ approaches unity.

The above argument is supported by the following analyses of the spin susceptibilities $\chi_{Co}^{spin}$ of the Co moments. Figure 7(a) shows $1/\chi_{Co}^{spin}$ of $(Pr_{1-y}Sm_y)_{0.7}Ca_{0.3}CoO_3$ measured with the external field $H = 0.1$ T under the zero-field cooling condition. The values of $\chi_{Co}^{spin}$ have been estimated by subtracting the susceptibility components, $\chi_{Pr}$ and $\chi_{Sm}$ contributed from $Pr^{3+}$ and $Sm^{3+}$, respectively. The $T$ independent contribution $\chi_0(y)$ has also been subtracted from the raw data. In the estimation of $\chi_{Pr}$, we have first calculated the number of $Co^{3+}$ ions excited to the IS state ($s=1$) in $PrCoO_3$ by using the $\delta E$ values of 1100 K and by considering the orbital degeneracy of the IS state (=18). (The LS ground state is not degenerated.) The $\delta E$ value can roughly be deduced from the $T$-dependence of the NMR longitudinal relaxation rate $1/T_1$.[21] Because $\chi_{Pr}$ is not large in the present $T$ region, the ambiguity of $\delta E$ does not bring any serious effect on the results. The $\chi_{Sm}$ value has been estimated by using the data of $SmScO_3$. To estimate $\chi_0(y)$, we have used the following method. At low temperatures, both $Co^{3+}$ and $Co^{4+}$ ions are in the LS ground state. So the slope of $1/\chi_{Co}^{spin} -T$ curve is expected, as $T$ decreases, to tend to the thick solid line in Fig. 7(a), which is described by the Curie-Weiss law with a Curie constant expected for the $s=1/2$ spins of $Co^{4+}$ ions. ($Co^{3+}$ ions, which are in the LS ground state, do not contribute to the susceptibility.) The $\chi_0(y)$ values are chosen to adjust the slopes of the $1/\chi_{Co}^{spin} -T$ curves in the $T$ region between 20 K ~ 60 K to the slope of the solid line. We note that the values of $\chi_0(y)$ used here roughly follows the relation $\chi_0(y)=-0.002+0.003y$ (emu/mol), though the relation may not have any significant physical meaning.

For $(Pr_{1-y}Sm_y)_{0.7}Ca_{0.3}CoO_3$, the energies $\delta E$ of the IS state of $Co^{3+}$ ($Co^{4+}$) ions with respect to the LS ground level of $Co^{3+}$ ($Co^{4+}$) ions can be determined by fitting the calculated $T$ dependence of $1/\chi_{Co}^{spin}$ to the experimentally obtained $1/\chi_{Co}^{spin}$-$T$ curve in the narrow $T$ region just above $T_s$ with $\delta E$ being the fitting parameter (The reason why the $T$ range is limited to the narrow region is shown later.). In this calculation of $\chi_{Co}^{spin}$ of $(Pr_{1-y}Sm_y)_{0.7}Ca_{0.3}CoO_3$ above $T_s$, the following method has been used. For the spin susceptibility of $Co^{3+}$ ions in the IS state with $s=1$ is proportional to $[(1-x)N_0g^2\mu_B^2s(s+1)/\{T-T_C(y)\}]\times[z\exp(-\delta E/k_BT)/\{1+z\exp(-\delta E/k_BT)\}]$, where $N_0$, $g$ and $z$ are the total number density of the Co ions, the electron $g$ factor (~2) and the degeneracy of the IS state (=18), respectively. For the spin susceptibility of $Co^{4+}$, similar calculations are also carried out for the LS state with $s=1/2$, $\delta E=0$ and $z=6$ and for the IS state with $s=3/2$ and $z=24$ (The $\delta E$ value of the IS state of $Co^{4+}$ is simply assumed to be equal to that of $Co^{3+}$.). We have not considered the effects of the HS state to the spin susceptibility, because they are expected to be very small.[17] Common values of $T_C(y)$ are used for both $Co^{3+}$ and $Co^{4+}$ ions, though it is not appropriate to accurately describe the observed $T$ dependence of $\chi_{Co}^{spin}$ by the use of $T_C(y)$ in the wide $T$ region, because $T_C(y)$ depends on the number density of the Co spins, which itself is $T$ dependent for the present system. It is the reason why the fitting is limited to the narrow region of $T$ just above $T_s$. The values of $T_C$ used here for $y=0.0$ is 60 K, and 15 K for $y=1.0$ and just the linear $y$ dependence of $T_C$ is assumed. The spin susceptibility of $\chi_{Co}^{spin}$ can be obtained as the sum of all the Co spins.

The thin solid lines in Fig. 7(a) show the results of the fittings carried out just above $T_s$. The obtained $\delta E$ values are shown in Fig. 7(b). Although our calculation of $\delta E$ has been done with rather crude assumptions, the $y$ dependent characteristics of $\delta E$ is qualitatively reproduced. In the region of the large $y$, the increase of $1/\chi_{Co}^{spin}$ found in the relatively high $T$ region of Fig. 7(a) with decreasing $T$, can be roughly understood, too, by the above calculation. We estimate here the number of Co ions excited to the IS state (or the electron number in the $e_g$ orbitals) at $T_s+0$, $n_{IS}$, by using the $\delta E$ values in Fig. 7(b) and by considering the orbital degeneracy of the levels. The results are shown in Fig. 7(c), where the solid line represents the number of $Pr^{3+}$ ions per formula unit. In the region of the relatively large $y$, $n_{IS}$ decreases with increasing $y$. It is smaller than the number of $Pr^{3+}$ ions. Due to this decrease of $n_{IS}$, the transition seems to gradually lose the cooperative character and becomes the (IS→LS) crossover-like, as $y$ approaches unity.

We also have studied the $x$-dependence of the transport and magnetic properties for the samples of $(Pr_{1-y}Sm_y)_{1-x}Ca_xCoO_3$ ($y=0.2$ and $0.3$; $0.0\leq x\leq 0.5$). The Ca doping introduces the holes into the Co sites. The electrical resistivities $\rho$ and the magnetic susceptibilities $\chi$ for the samples of $(Pr_{0.7}Sm_{0.3})_{1-x}Ca_xCoO_3$ are plotted against $T$ in Figs. 8 and 9, respectively. The data of $\chi$ were taken under the condition of the zero field cooling. In the region of $x\geq 0.3$, $T_s$ slightly increases with increasing $x$. As $x$ decreases from 0.3, the transition becomes vague. Figure 10(a) shows the inverse spin susceptibilities of the Co moments, $1/\chi_{Co}^{spin}$ for $(Pr_{0.7}Sm_{0.3})_{1-x}Ca_xCoO_3$. The values of $\chi_{Co}^{spin}$ are estimated by subtracting the contributions from $Pr^{3+}$ and $Sm^{3+}$ and the $T$ independent one, whose magnitude is chosen to adjust the $x$ dependent Curie constant or the slope of the $1/\chi_{Co}^{spin}$ -$T$ curve in the low temperature region in the similar way to the case of Fig. 7(a). The thin solid lines in Fig. 10(a) represent the results of the fittings just above $T_s$ with $\delta E$ being the fitting parameter to the observed $1/\chi_{Co}^{spin}$-$T$ curves (The transition (or crossover) temperature $T_s$ is roughly defined to be the onset temperature of the increase of $1/\chi_{Co}^{spin}$ with decreasing $T$.). Even though the fittings have been carried out only in the narrow region of $T$ just above $T_s$, the calculations for the relatively small $x$ seem to describe the qualitative $T$ dependence of $1/\chi_{Co}^{spin}$ in the rather wide $T$ region above $T_s$. It is due to the fact that the Curie-Weiss temperatures $T_C$, which appear in the calculation, are much smaller than the relevant temperatures ($>T_s$).

The electronic energies of the IS state of $Co^{3+}$ ($Co^{4+}$) in $(Pr_{0.7}Sm_{0.3})_{1-x}Ca_xCoO_3$ obtained by the fittings are plotted against $x$ in Fig. 10(b). It increases with decreasing $x$, which is consistent with the results of our previous paper.[17] In the region of the small $x$ values, the increase of $1/\chi_{Co}^{spin}$ with decreasing $T$ found in the high $T$ region seems to be understood without the phase transition. We have estimated $n_{IS}$ at $T_s+0$ by using the $\delta E$ values in Fig. 10(b), and the results are in Fig. 10(c). The $n_{IS}$ value decreases with decreasing $x$ and as $n_{IS}$ decreases, the transition becomes the crossover-like one, losing the cooperative nature, as in the case shown in Fig. 6.

Figure 11 summarizes the $T$-$x$ phase diagram for $y=0.2$ and $0.3$. The gradual change of the nature of the IS→LS transition is depicted schematically by the broadening of the shaded $T$ region. The increase of $T_s$ with decreasing $x$ is due to the increase of $\delta E$, which stabilizes the LS state, while the increase of $T_s$ with increasing $x$ in the region of $x >0.3$ is due to the increase of $n_{IS}$, which enhances the cooperative nature of the IS→LS change.

We have so far presented how the nature of the spin state change depends on $x$ and $y$. There still remains, however, a question if the transition exists even in systems without Pr and/or Ca. To answer the question, we have been studying the transport behavior of $R_{1-x}A_xCoO_3$ (R=La, Pr, $La_{0.2}Nd_{0.8}$ and Nd: A=Ba, Sr and Ca) up to high pressure $p$. In the previous papers,[17, 21] we pointed out that the transition was observed only for (R, A) = (Pr, Ca) system. Furthermore, in the present work, we have searched for the transition again in the sample of $Nd_{1-x}Ca_xCoO_3$ with $x = 0.3$ and $0.4$ under high pressure up to 100 kbar. Figure 12 shows the electrical resistivities ρ of the samples of $Nd_{0.6}Ca_{0.4}CoO_3$ under the various values of the applied pressure $p$. The resistivity ρ increases with increasing $p$,[17, 21, 30] due to the decrease of the electron number in the $e_g$ orbitals through the increase of $\Delta_c$ and $\delta E$ with increasing $p$. This system does not exhibit the phase transition up to 100 kbar. If the transition mechanism commonly exists in $R_{1-x}Ca_xCoO_3$ with various species of R, the application of such the high pressure may induce the transition in $Nd_{1-x}Ca_xCoO_3$, as in the case of $Pr_{1-x}Ca_xCoO_3$, though the LS state is expected to be more stable in the system of (Nd, Ca) than in the system of (Pr, Ca), because the ionic radius of $Nd^{3+}$ is slightly smaller than that of $Pr^{3+}$. We have extended the studies to the system of $(Nd_{1-y}Tb_y)_{0.7}Ca_{0.3}CoO_3$ ($y=$ 0, 0.1 and 0.2) at ambient pressure, where we have not found any indication of the phase transition, while in the system of $(Pr_{1-y}Tb_y)_{0.7}Ca_{0.3}CoO_3$, the transition is induced by the small amount (10%) of the Pr substitution with Tb.[17] One might think that the non-existence of the transition is due to the small $n_{IS}$. However, estimating the spin susceptibilities of the Co moments, $\chi_{Co}^{spin}$ of $Nd_{0.7}Ca_{0.3}CoO_3$, the value of $n_{IS}$ is not so small that the system does not exhibit the phase transition: The value of $\chi_{Co}^{spin}$ of $Nd_{0.7}Ca_{0.3}CoO_3$ are estimated by subtracting the susceptibility components, $\chi_{Nd}$ of $Nd^{3+}$ moments, and the $T$ independent term, as has been done in the above analyses for $Pr_{0.7}Ca_{0.3}CoO_3$. The $n_{IS}$ value found there for $Nd_{0.7}Ca_{0.3}CoO_3$ is nearly equal to the one of the sample of $(Pr_{0.6}Sm_{0.4})_{0.7}Ca_{0.3}CoO_3$, which exhibits a clear transition (see Fig. 7(a)). These results indicate that, for the occurrence of the phase transition, the existence of Pr or Pr and Ca seems to be necessary.

Now, we consider the structural characteristics of $Pr_{1-x}Ca_xCoO_3$. There are twelve O atoms surrounding a Pr atom. Among these oxygen atoms, three are very close to the Pr atom (bond length of ~(2.35-2.37) Å,[17] which should be compared with the length of ~(2.45-2.47) Å found in $PrBa_2Cu_3O_7$.[31]) Furthermore, at $T_s$, they decrease abruptly with decreasing $T$. Recently, Nishibori et al.[32] have studied the charge density distribution of a sample with $x=0.3$ by the maximum entropy method (MEM), using the powder X-ray diffraction data taken in SPring-8 and observed the experimental evidence for the Pr-O hybridization below $T_s$ for these three bonds. The result is similar to the Pr 4f-O 2p$\pi$ hybridization in $PrBa_2Cu_3O_7$,[33] which was also found to take place by the MEM analyses.[34] It is interesting to consider effects of the Pr 4f-O 2p$\pi$ hybridization as the origin of the first order phase transition accompanied by the spin state transition in $Pr_{1-x}Ca_xCoO_3$.

## 4. Conclusion

The transport, thermal and magnetic properties of $(Pr_{1-y}Sm_y)_{1-x}Ca_xCoO_3$ have been studied and the $T$-$y$ and $T$-$x$ phase diagrams have been constructed. By analyzing the magnetic susceptibility, the number of Co ions in the IS state (or the electron number in the $e_g$ orbitals), $n_{IS}$, are roughly estimated at $T_s+0$. As $y$ approaches unity or $x$ approaches zero, $n_{IS}$ becomes small. As the results of this $n_{IS}$ reduction, the IS to LS change with decreasing $T$ loses the cooperative nature and gradually becomes the (IS→LS) crossover-like one. We have not found any experimental evidence that indicates the existence of the first order phase transition in the perovskite Co oxides without (Pr, Ca), which implies that the Pr 4f-O 2pπ hybridization may be relevant to the occurrence of the transition.

Acknowledgements –The authors thank Prof. M. Sakata and Dr. E. Nishibori of Nagoya University for their discussion. The synchrotron radiation experiments were performed at the BL10XU in the SPring–8 with the approval of the Japan Synchrotron Radiation Research Institute. (Proposal No. 2003A-5004-LD-np). The work is supported by Grants-in-Aid for Scientific Research from the Japan Society for the Promotion of Science (JSPS) and by Grants-in-Aid on priority area from the Ministry of Education, Culture, Sports, Science and Technology.

Figure captions

Fig. 1   Unit cell volume of $(Pr_{1-y}Sm_y)_{0.7}Ca_{0.3}CoO_3$ is plotted against $y$.

Fig. 2   Temperature dependence of the electrical resistivities $\rho$ of $(Pr_{1-y}Sm_y)_{0.7}Ca_{0.3}CoO_3$ is shown for various values of $y$.

Fig. 3   Temperature dependence of the magnetic susceptibilities $\chi$ of $(Pr_{1-y}Sm_y)_{0.7}Ca_{0.3}CoO_3$ measured with the magnetic field $H=0.1$ T is shown for various values of $y$. The data were taken under the zero-field-cooling condition.

Fig. 4   Thermoelectric powers $S$ of $(Pr_{1-y}Sm_y)_{0.7}Ca_{0.3}CoO_3$ are plotted against $T$ for various values of $y$.

Fig. 5   Temperature dependence of the specific heat $C$ of $(Pr_{1-y}Sm_y)_{0.7}Ca_{0.3}CoO_3$ is shown for various values of $y$. The inset shows $C/T$ of $(Pr_{1-y}Sm_y)_{0.7}Ca_{0.3}CoO_3$ against $T$ for several $y$ values.

Fig. 6   Schematic $T$-$y$ phase diagram of $(Pr_{1-y}Sm_y)_{0.7}Ca_{0.3}CoO_3$. The structural transition temperature $T_s$ is broadened as $y$ approaches unity and becomes crossover-like.

Fig. 7   (a) Inverse spin susceptibilities of the Co moments, $1/\chi_{Co}^{spin}$, of $(Pr_{1-y}Sm_y)_{0.7}Ca_{0.3}CoO_3$ measured under the zero-field cooling condition with the magnetic field $H=0.1$ T are shown against $T$. The values of $\chi_{Co}^{spin}$ have been estimated by subtracting the susceptibility components, $\chi_{Pr}$ and $\chi_{Sm}$ of $Pr^{3+}$ and $Sm^{3+}$, respectively. The $T$ independent contribution $\chi_0(y)$ has also been subtracted from the raw data. The thick solid line represents the Curie-Weiss law with a Curie constant expected for the $s=1/2$ spins of $Co^{4+}$ ions. The thin solid lines show the results calculated for the parameters of $\delta E$ values shown in Fig. 7(b). See text for details. (b) The difference of the electronic energies, $\delta E$ between the IS and LS states of $Co^{3+}$ and $Co^{4+}$ ions in $(Pr_{1-y}Sm_y)_{0.7}Ca_{0.3}CoO_3$ are shown against $y$. They are estimated just above $T_s$ by fitting the calculated $\chi_{Co}^{spin}$-$T$ curves to the observed ones. See text for details. (c) The number of Co ions excited to the IS state at $T_s+0$, $n_{IS}$, is plotted against $y$. They were calculated by using $\delta E$ values in Fig. 7(b) and by considering the orbital degeneracy of the LS and IS states. Solid line represents the number of $Pr^{3+}$ ions per a formula unit.

Fig. 8   Temperature dependence of the electrical resistivities $\rho$ of $(Pr_{0.7}Sm_{0.3})_{1-x}Ca_xCoO_3$ is shown for various values of $x$.

Fig. 9   Temperature dependence of the magnetic susceptibilities $\chi$ of $(Pr_{0.7}Sm_{0.3})_{1-x}Ca_xCoO_3$ measured with the magnetic field $H=0.1$ T is shown for various values of $x$. The data were taken under the zero-field-cooling condition.

Fig.10 (a) Inverse spin susceptibilities of the Co moments, $1/\chi_{Co}^{spin}$, of $(Pr_{0.7}Sm_{0.3})_{1-x}Ca_xCoO_3$ measured under the zero-field cooling condition with the magnetic field $H=0.1$ T are shown against $T$ for various $x$ values. The values of $\chi_{Co}^{spin}$ are estimated by subtracting the susceptibility of $Pr^{3+}$ and $Sm^{3+}$ and the $T$ independent susceptibility $\chi_0(y)$ from the raw data. The thin solid lines show the results calculated for the parameters of $\delta E$ values shown in Fig. 10(b). See text for details. (b) The differences of the electronic energies, $\delta E$ between the IS and LS states of $Co^{3+}$ and $Co^{4+}$ ions in $(Pr_{0.7}Sm_{0.3})_{1-x}Ca_xCoO_3$ are shown against $x$. They are estimated by fitting the calculated $\chi_{Co}^{spin}$-$T$ curves to the observed ones. See text for details. (c) Numbers of Co ions excited to the IS state at $T_s+0$, $n_{IS}$, in $(Pr_{0.7}Sm_{0.3})_{1-x}Ca_xCoO_3$ are shown for various values of $x$. They were calculated by using $\delta E$ values in Fig. 10(b) and by considering the orbital degeneracy of the LS and IS states.

Fig. 11 Schematic $T$-$x$ phase diagram of $(Pr_{1-y}Sm_y)_{1-x}Ca_xCoO_3$ ($y=0.2$ and 0.3). $T_s$ is broadened as $x$ approaches 0 and the transition becomes crossover-like.

Fig. 12 Electrical resistivities $\rho(p,T)/\rho(1bar,300K)$ of $Nd_{0.6}Ca_{0.4}CoO_3$ taken under various values of applied pressure $p$ are shown against $T$.

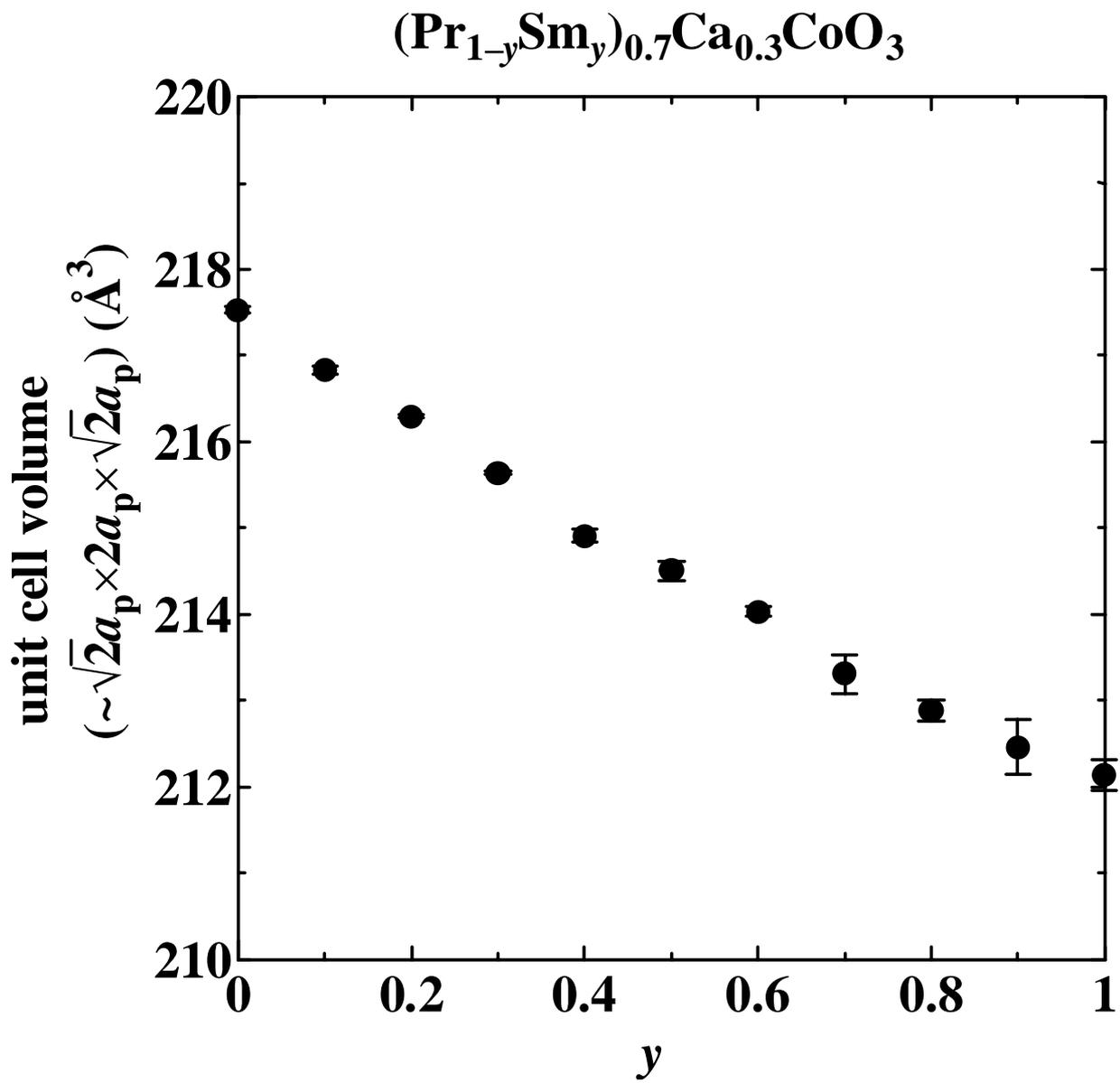

Fig.1
Fujita et al.

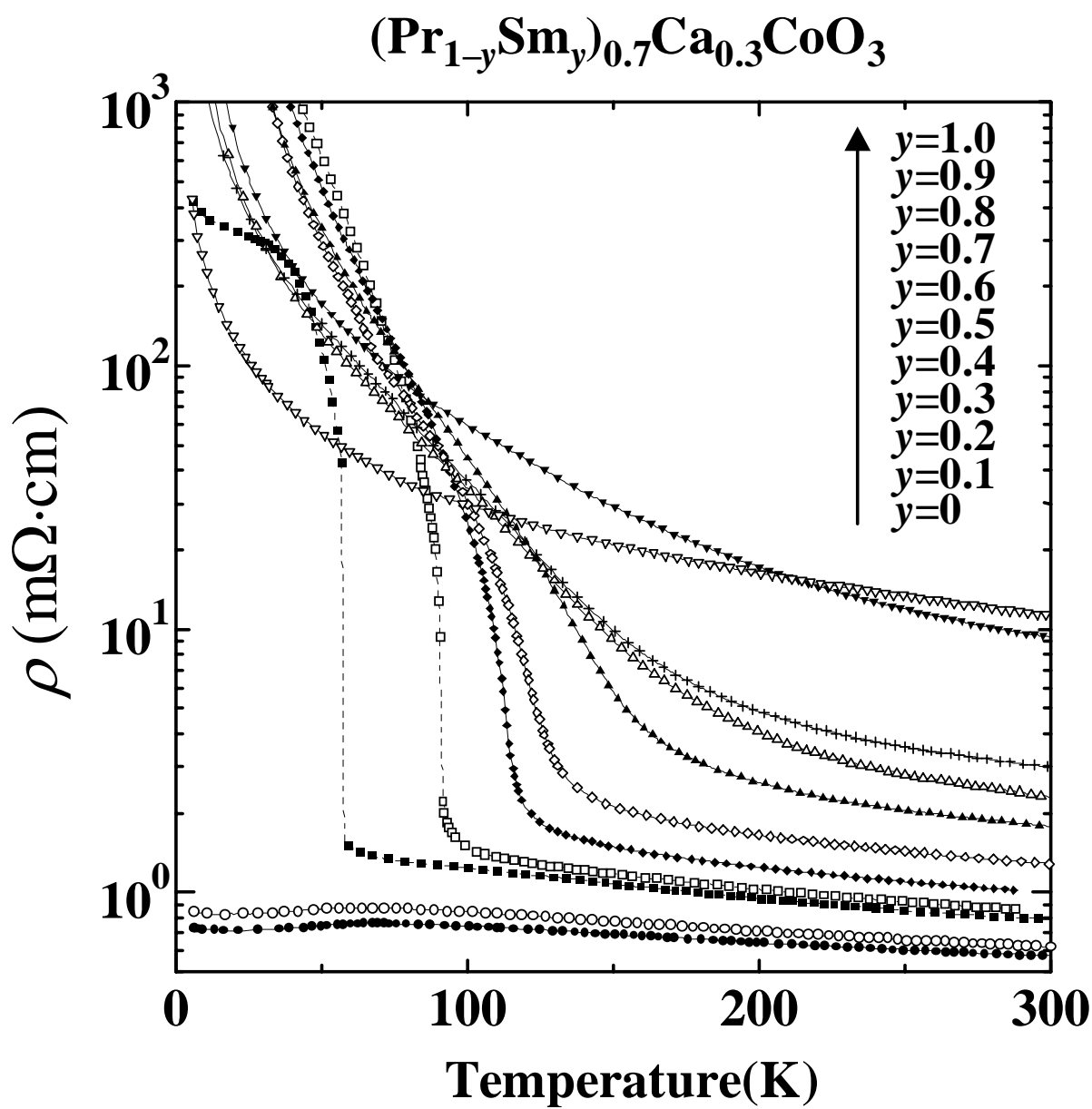

Fig.2
Fujita et al.

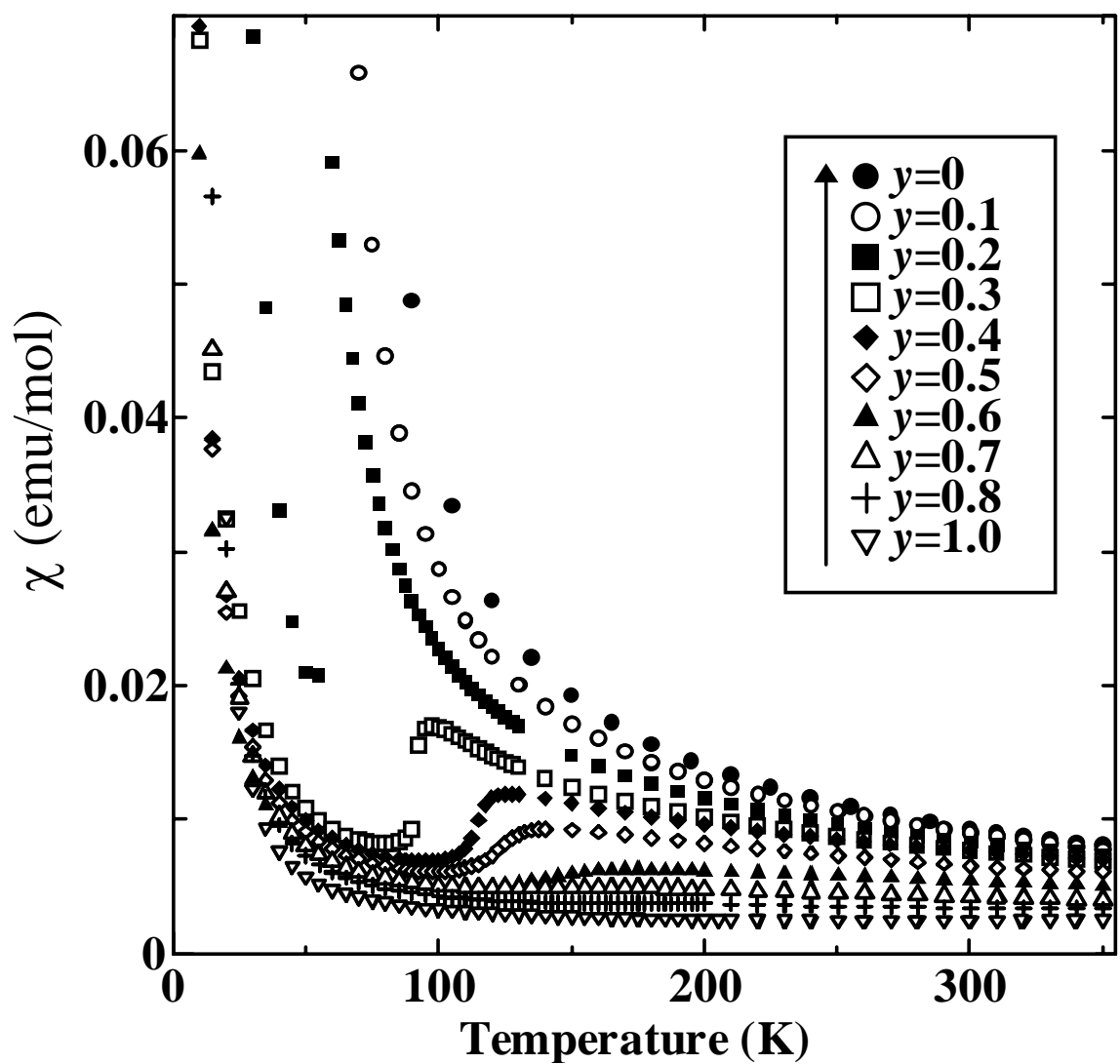

Fig.3
Fujita et al.

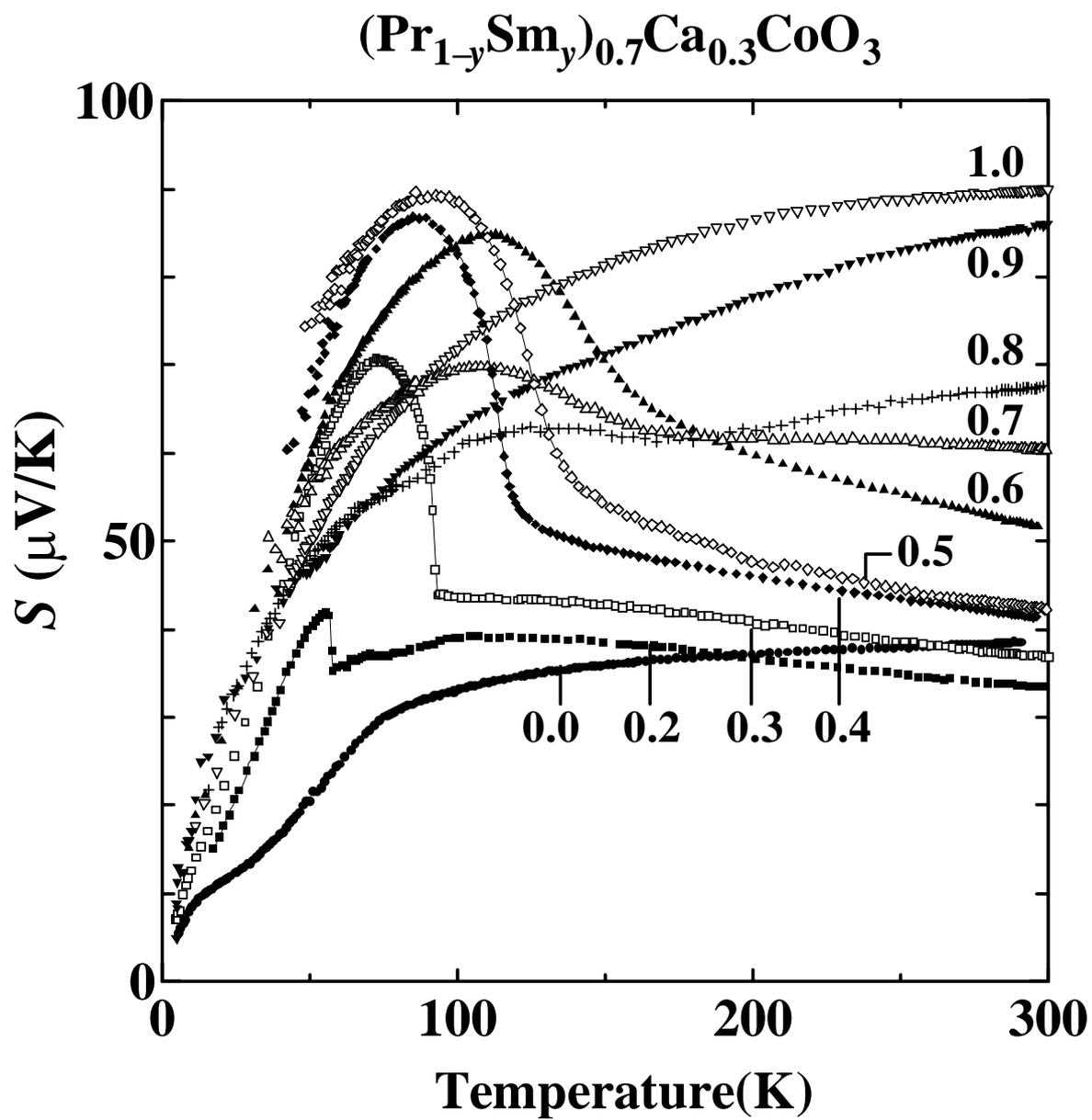

Fig.4
Fujita *et al.*

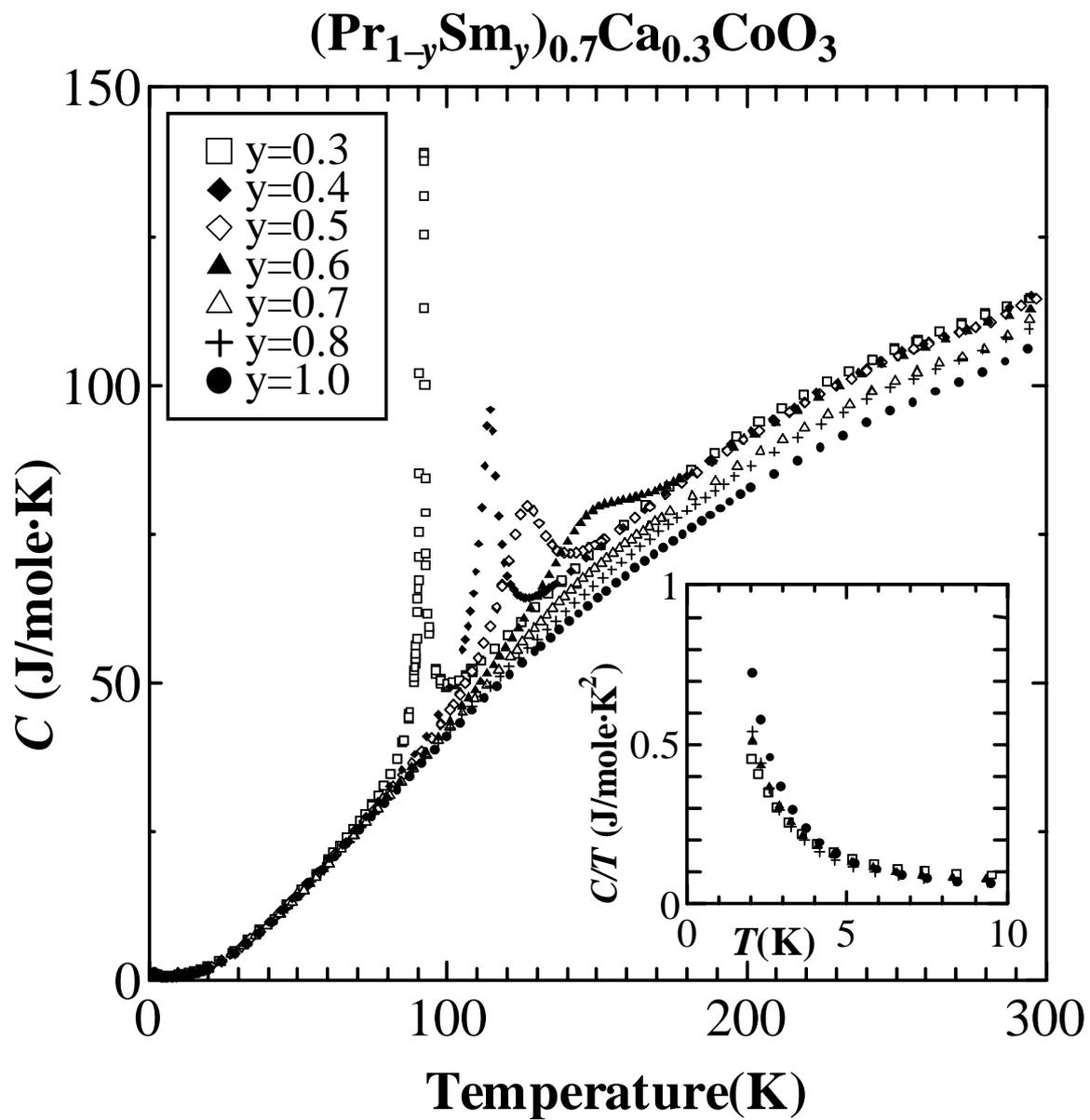

Fig.5
Fujita et al.

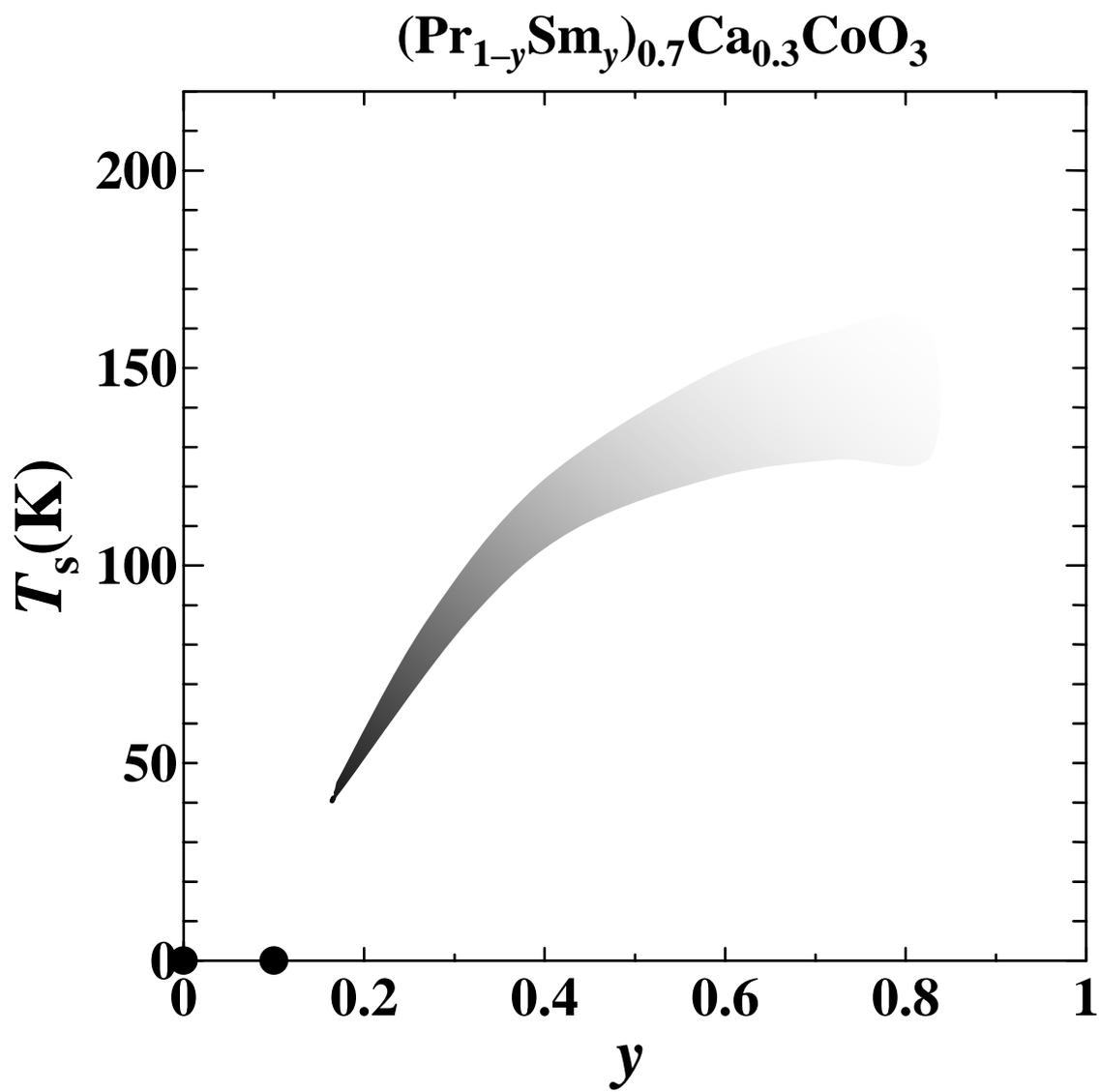

Fig.6
Fujita et al.

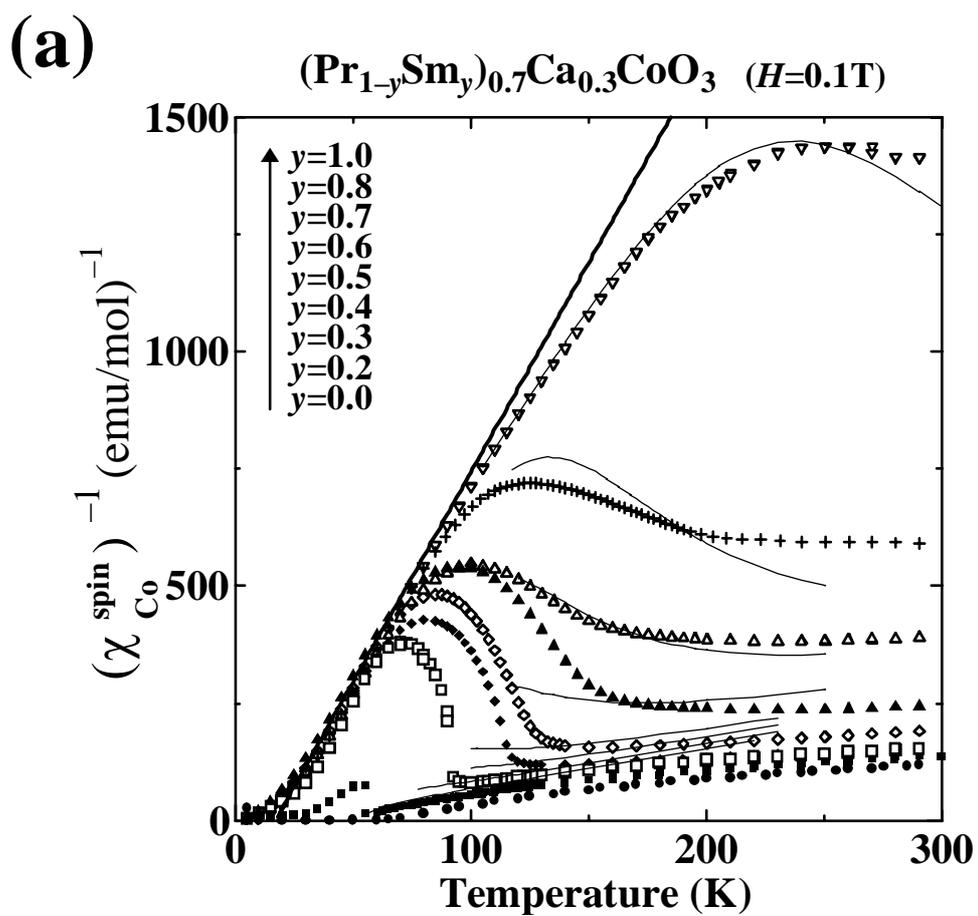

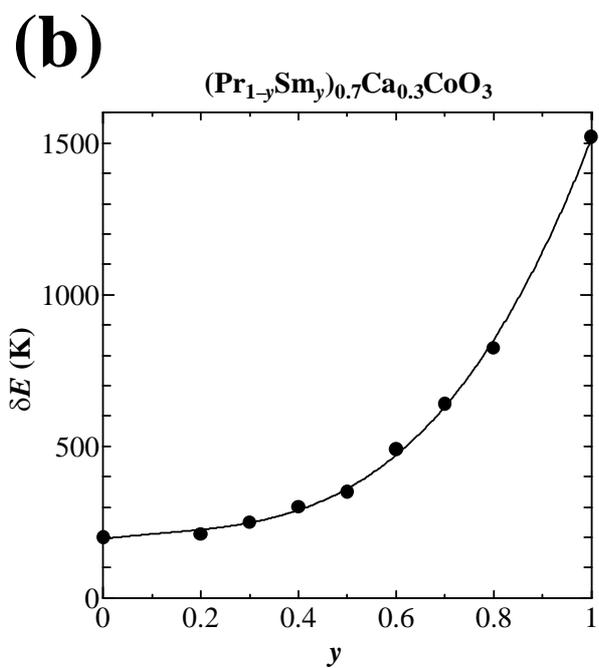
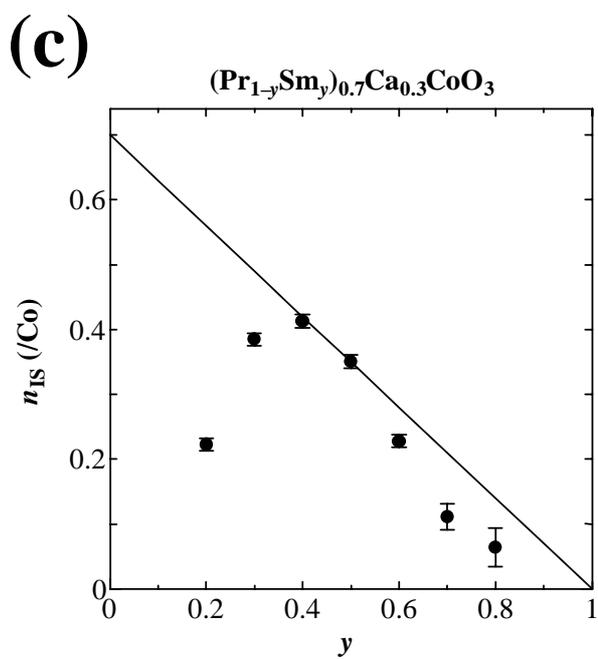

Fig.7 Fujita et al.

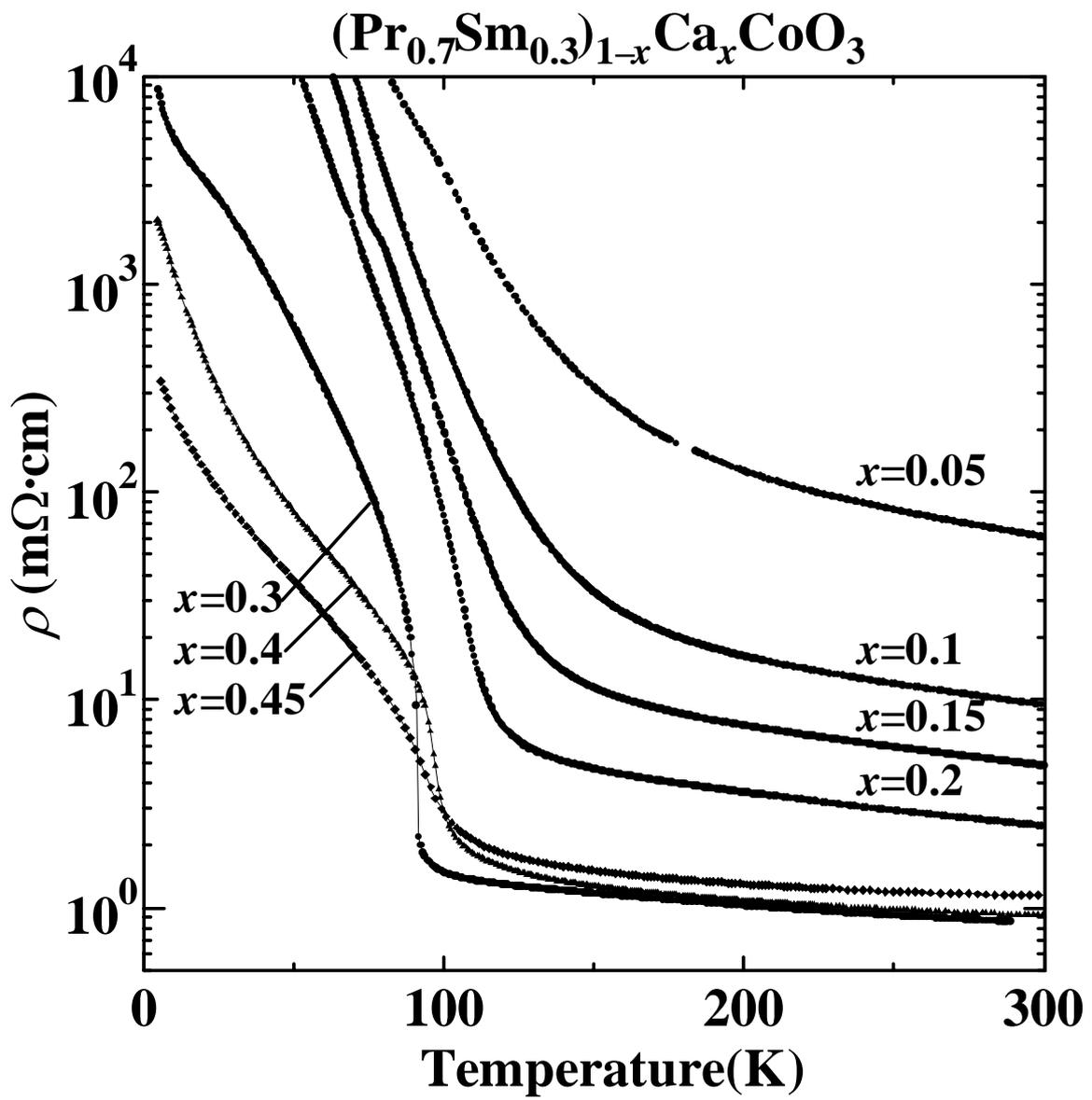

Fig.8
Fujita *et al.*

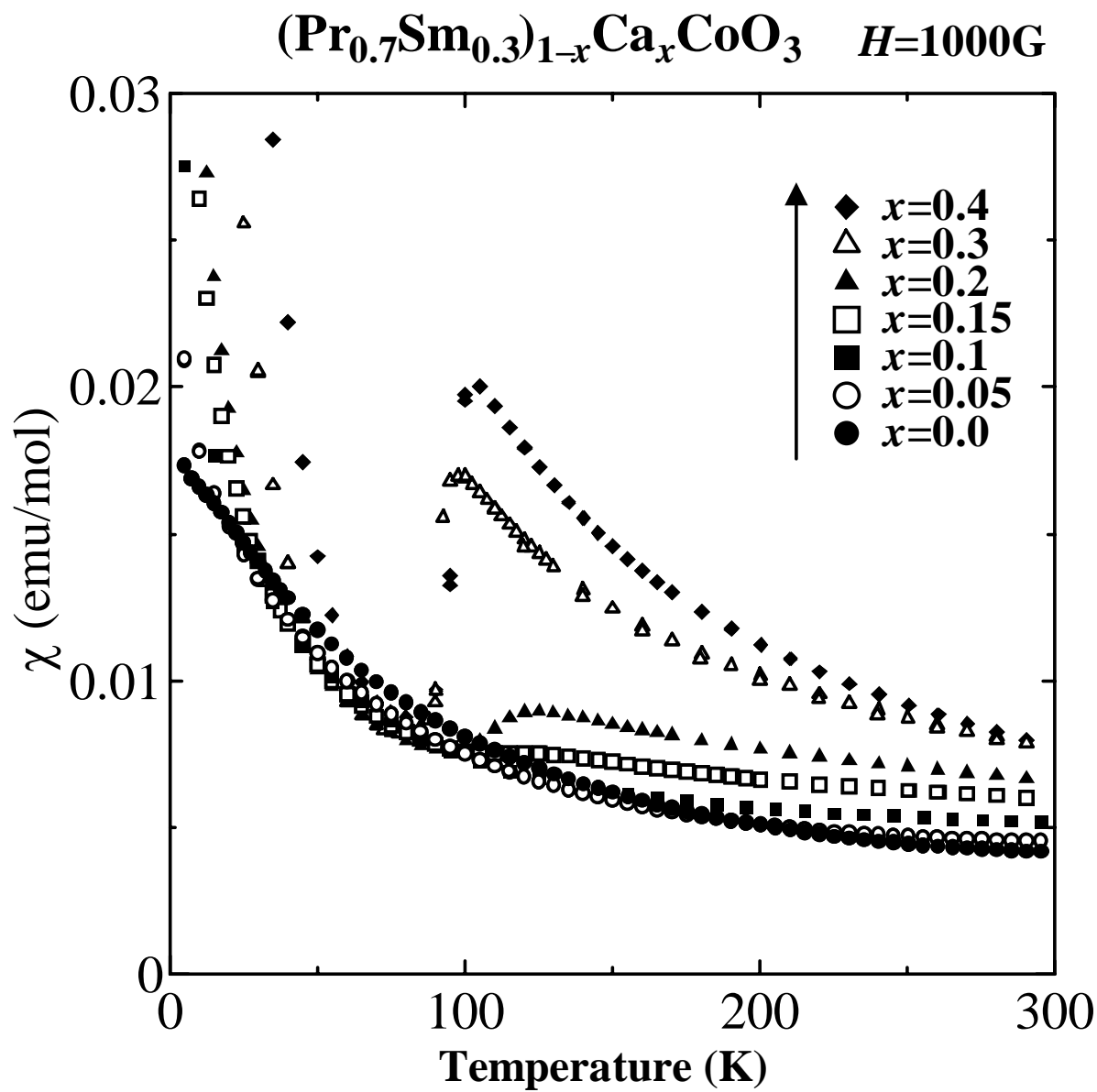

Fig.9
Fujita *et al.*

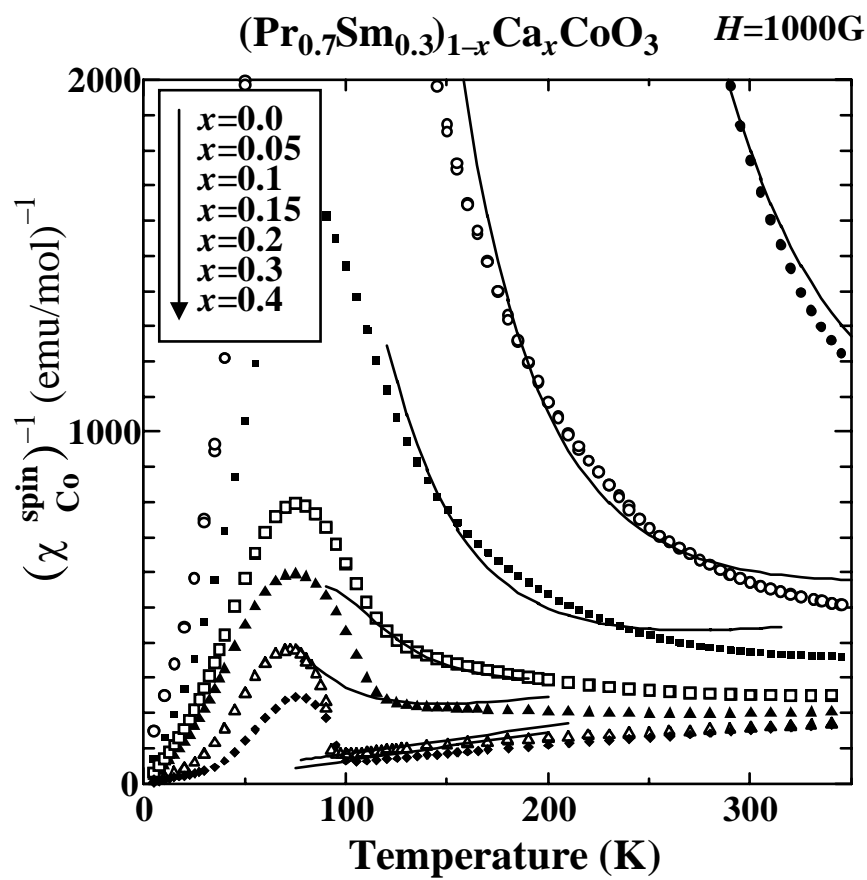

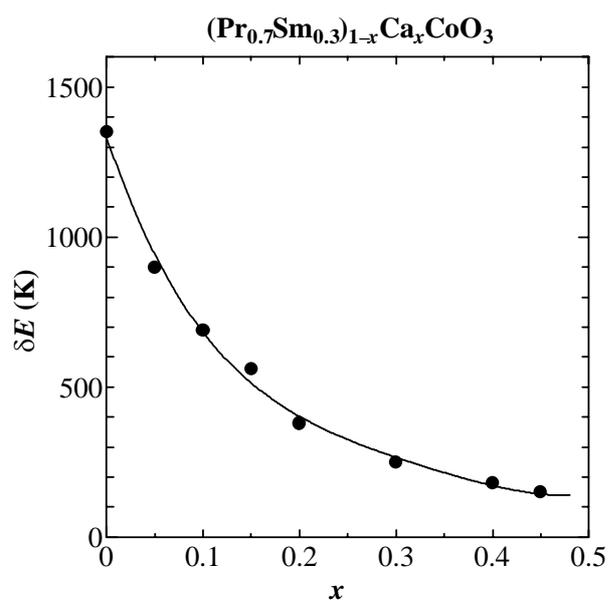

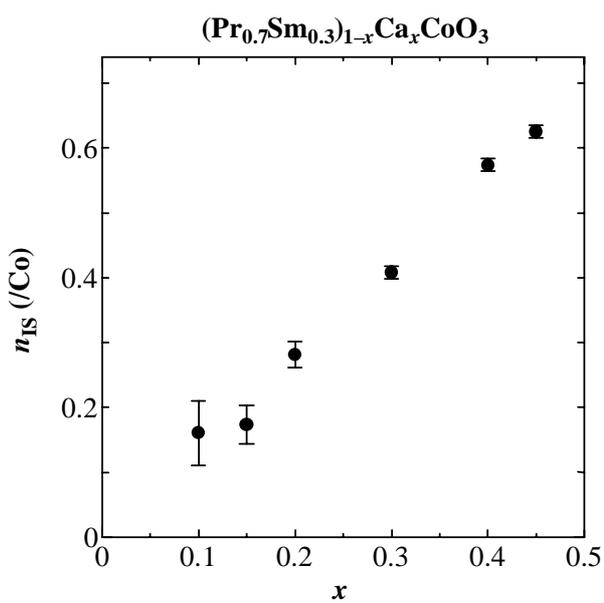

Fig.10
Fujita *et al.*

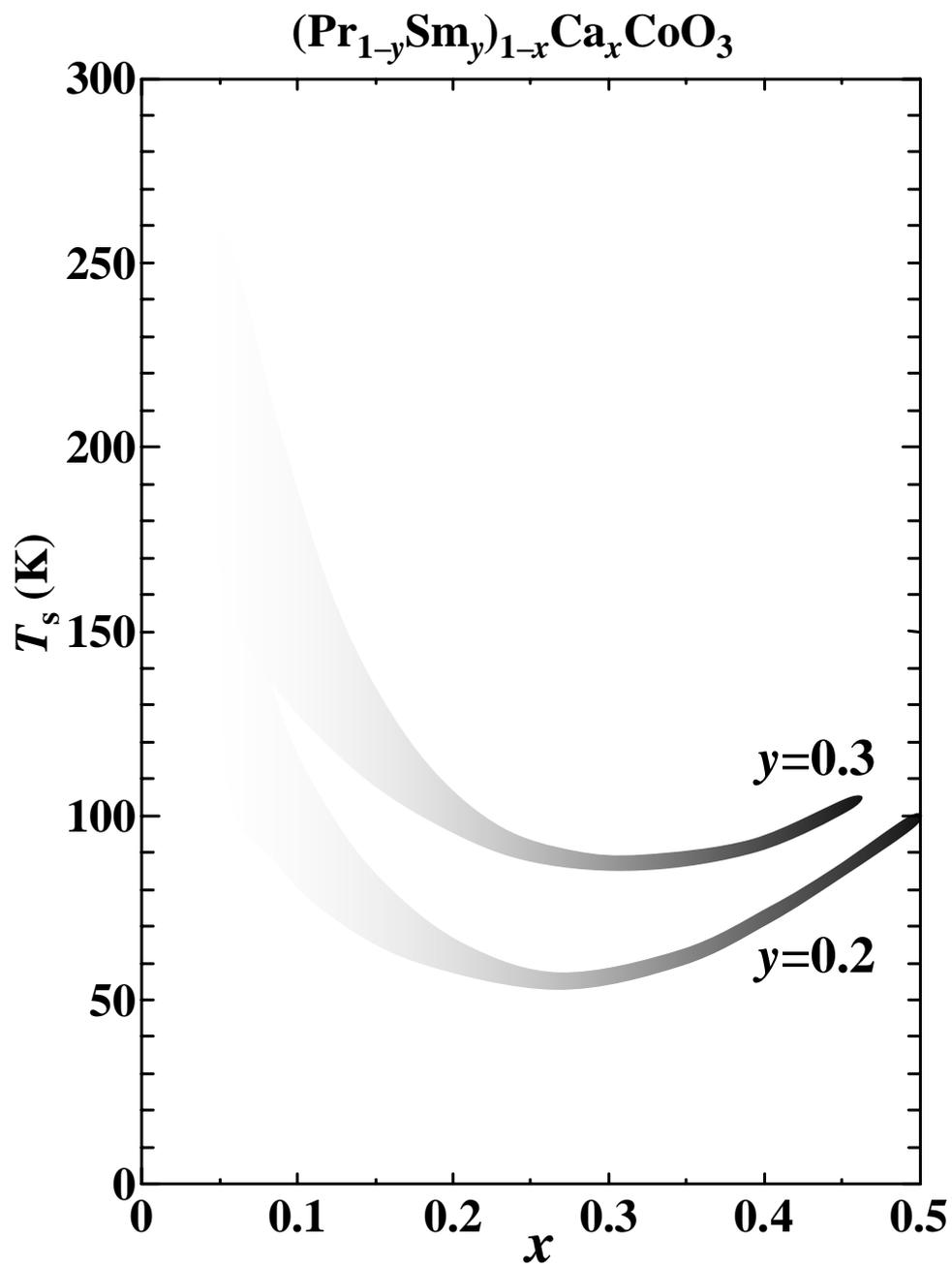

Fig.11
Fujita *et al.*

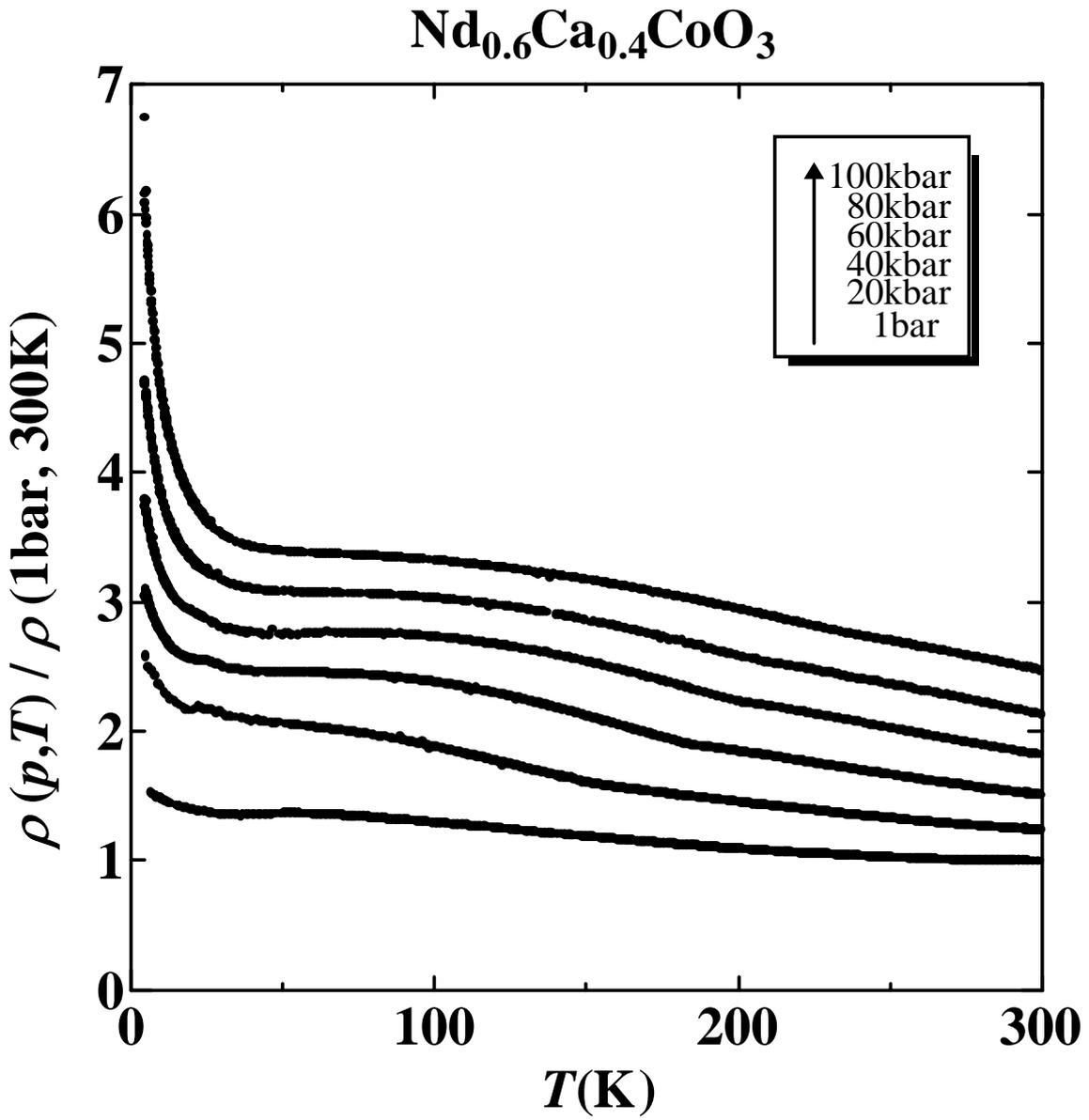

Fig.12
Fujita *et al.*